\documentstyle[12pt]{article}

\newcommand{\be}{\begin{equation}}
\newcommand{\ee}{\end{equation}}
\newcommand{\bea}{\begin{eqnarray}}
\newcommand{\eea}{\end{eqnarray}}

\begin{document}
\title{
\hfill \normalsize{KCL-TH-94-3} \\
\vskip2cm
\LARGE{On representations of super coalgebras}
}
\author{Andreas H\"uffmann \\
	Department of Mathematics \\
	King's College \\
	Strand \\
	London WC2R 2LS}
\date{}
\maketitle

\begin{abstract}
The general structure of the representation theory of a $Z_2$-graded
coalgebra is discussed. The result contains the structure of Fourier
analysis on compact supergroups and quantisations thereof as a special
case. The general linear supergroups serve as an explicit illustration
and the simplest example is carried out in detail.
\end{abstract}

\newpage

\parindent15pt
\jot10pt

\newtheorem{lemma}{Lemma}
\newtheorem{corollary}{Corollary}
\newtheorem{proposition}{Proposition}
\newtheorem{theorem}{Theorem}

\section{Introduction}
Although the theory of Fourier analysis on supergroups or related coset
supermanifolds has received considerable attention the results are still far
from a systematic theory that extends the classical theory for Lie groups
and symmetric spaces. Recent progress concerning spherical Fourier analysis on
certain noncompact Riemannian super coset spaces has been made by Zirnbauer
\cite{Zir:93} \cite{Zir:90a}. His work was directly motivated by the demands
of supersymmetric field theoretical models for an electron moving in a
disordered environment \cite{Zir:90b} \cite{Zir:92}. The peculiarities of
these theories stimulated the wish for a better understanding of compact
supergroups as well. These appear for instance as angle variables upon the
introduction of polar coordinates in supersymmetric models.

A systematic study of compact unitary supergroups finally led to the
general setting that is to be presented in this paper. It is basically a
structure theorem for graded coalgebras with impact not only on super groups
but as well on super quantum groups or graded Hopf algebras respectively.
The arguments are simple and were finally settled in the foundations on the
level of Zorn's lemma. After this work had been completed, Green's treatment
of locally finite representations came to my attention \cite{Gre:76}. It
contains the structure theorem, derived by different arguments without
exploiting the possibility of a grading.

The text falls into three major units. At first the general structure of
graded coalgebras and comodules is discussed. This reveals the basic
structure of Fourier analysis of graded coalgebras. The role of induced
re\-presentations is outlined. Second, the coalgebra corresponding to
locally finite modules of the general linear supergroup is
analysed; Kac modules, Kac filtrations and their role for the structure of
its representations are discussed. Finally the general theory is illustrated
by the simplest examples. Here a coalgebra related to the Lie superalgebra
$gl(1,1;{\bf C})$ is considered in detail. In spite of its very simple nature
almost all general phenomena can be observed with a minimal computational
effort. Contact is made with existing literature on $gl(2,1;{\bf C})$ and a
summary is given.

Although the results of Section 2 extend further, the convention that the
term 'graded' will always mean '$Z_2$-graded' is adopted as a framework.
Since all objects come in a graded context the explicit annotation of
'graded' will be dropped unless in cases where some confusion may arise. For
example '(semi)simple' has the meaning of 'graded (semi)simple'. If a simple
comodule is required to be irreducible in the non graded sense this will be
emphasized by using 'absolutely simple'. Furthermore a basis is to be thought
of as a homogeneous basis of some comodule and homomorphisms are homogeneous
of degree zero. The terms 'comodule' and 'representation' will be used
synonymously.

Since the intention was not to extensively reproduce previously published
material on Lie superalgebras, the text is kept informal at some stages. For
instance root systems and weight bases are not discussed in detail and
reference to the literature is given whenever these notions are needed to put
a subject into a wider perspective. However, the basic line of thought is
presented self contained and should be readable without in depth knowledge of
the vast literature on supergroups and -algebras.

\section{Representations of graded coalgebras}

At first, the most basic consequences from the definition of a coalgebra $C$
are to be recalled. These follow directly from its two structure
homomorphisms, that is a coassociative comultiplication
$\Delta :C \to C \otimes C$ and a counit $\varepsilon :C \to k$, which is a
mapping into the base field $k$. To be explicit, coassociativity means
$(\Delta \otimes id)\Delta = (id \otimes \Delta )\Delta $ and the counit
satisfies
$\cdot (\varepsilon \otimes id)\Delta
= \cdot (id \otimes \varepsilon )\Delta = id$,
$\cdot$ being the multiplication with scalars.
$C$ is considered as a (right or left) comodule over itself, with
structure map
$\Delta : C \to C \otimes C$.

The coassociativity is responsible for the
coalgebra being locally finite dimensional \cite{Wat:79}. That means, if
$x \in C$ is arbitrary but fixed, then there is a finite dimensional
subcomodule $V$ such that $x \in V \subset C$. The simple proof is as follows.
Let $\lbrace b_j \rbrace_{j \in J} \subset C$ be a basis,
$\Delta(b_j) = \sum_{k \in J} c_{jk} \otimes b_{k}$. Fix $x \in C$. Write
$\Delta(x) = \sum_{j \in J_x} c_j \otimes b_j$ with $J_x \subset J$
finite. Coassociativity means
$\sum_{j \in J_x} \Delta(c_j) \otimes b_j =
(\Delta \otimes id)\Delta(x) = (id \otimes \Delta)\Delta(x)
= \sum_{k \in J} (\sum_{j \in J_x}  c_j \otimes c_{jk}) \otimes b_{k}$.
Hence the span of $\lbrace c_j \rbrace_{j \in J_x} \cup \lbrace x \rbrace$
is a finite dimensional subcomodule of $C$.

The existence of a counit implies that every subcomodule of $C$ is contained
in the span of its coefficients; in other words, a coalgebra is the direct
limit of finite coalgebras formed by the coefficients of its finite
dimensional comodules.
For that let $\lbrace v_i \rbrace_{i \in I}$ be a basis of a subcomodule
$V \subset C$ and $\Delta (v_i) = \sum_{j \in I} v_j \otimes a^j_i$. Then
$v_i = (\cdot(\varepsilon \otimes id)\Delta )(v_i)
= \sum_{j \in I} \varepsilon (v_j) a^j_i.$

The definition of a comodules structure map, say $\beta :V \to V \otimes C$,
enforces $V$ to be included in a direct sum of copies of $C$.
Injectivity is immediate from the left inverse of $\beta $,
$\cdot (id \otimes \varepsilon ) \beta = id$. The fact that $\beta $ is a
comodule map is nothing but
$(\beta \otimes id)\beta = (id \otimes \Delta)\beta $ where
$id \otimes \Delta$ is defined as the comodule structure on
$V \otimes C$. It should be noted that this is not a tensor product of
representations, which is not defined unless the coalgebra is endowed with
a compatible algebra structure. For a finite dimensional $V$,
$dim(V) = n$, it means an inclusion
$\beta:V \hookrightarrow C^n$ \cite{Wat:79}. Moreover, all
comodules are locally finite and the problem of finding all finite
dimensional representations has as its first step the question of how to
decompose $C$.

To make contact to the physical terminology one should imagine $C$ as some
sort of functions on a quantum semigroup. Passing to a quantum group in
general requires enlarging $C$. This has to be done since the existence of
an antipode demands the invertibility of matrices,  which might be
impossible within $C$ itself. The natural approach seems to be Manin's
construction of Hopf envelopes \cite{Man:91}. The fundamental problem is the
Fourier analysis of $C$, which may drastically change in its nature with the
passage to a Hopf envelope. This will appear explicitly in the discussion of
the general linear Lie superalgebra.

{}From the point of view of category theory, attempts to study representations
lead to the investigation of injective objects. The structure theory of
coalgebras appears to be a direct generalisation of the representation
theory of finite groups. It is beautifully developed in Green's work on
locally finite representations \cite{Gre:76}. However, this aspect seems not
to have been considered within the mathematical physics literature. For
that a brief discussion is to be given. It is based on Zorn's lemma as an
alternative to the approach in the above reference.

For a $C$-comodule $W$ let $\sigma (W)$ be its socle, that is its semisimple
subcomodule. For every semisimple $V$ let $F_V$ be the family of comodules
the socles of which are $V$. $F_V$ has a partial ordering by inclusions.
The following lemma gives the basic intuition about the representation theory
of $C$. For its proof note that a homomorphism of comodules is injective if
and only if its restriction to the socle is.

\begin{lemma} \label{repint} \parindent0pt
Let $V$ be a semisimple $C$-comodule. The following assertions are equivalent
and hold for every coalgebra.

\begin{enumerate}
\item $F_V$ has a greatest element.
\item \label{2} Every totally ordered subset of $F_V$ has an upper bound in
$F_V$.
\item There is a maximal element in $F_V$.
\item \label{4} There is a $X \in F_V$ such that $X \subseteq W$ always
implies $W \simeq X \oplus X'$.
\item For every semisimple $X$ let $I_X \in F_X$ be a fixed choice of an
element satisfying \ref{4}. Then for every comodule
$W$ one has $W \hookrightarrow I_{\sigma (W)}$.
\end{enumerate}
\end{lemma}

{\it Proof}. $1. \Rightarrow 2.$ is trivial and $2. \Rightarrow 3.$ is
Zorn's lemma.

$3. \Rightarrow 4.$ Let $X \in F_V$ be a maximal element, $X \subseteq W$.
Let $X' \subset W$ be a maximal complementary subcomodule to $X$ in $W$.
Then $X'$ is also maximal complementary to $V \subseteq X \subseteq W$.
Hence $W/X' \in F_V$. But $X \subseteq W/X'$ via $x \mapsto x + X'$ and by
maximality $X = W/X'$. Thus $0 \to X' \to W \to W/X' \to 0$ splits and
$W \simeq X \oplus X'$.

$4. \Rightarrow 5.$ For a comodule $W$ let
$ \delta : \sigma (W) \to W \oplus I_{\sigma (W)}$ be given by
$x \mapsto \delta(x)=(x,x)$. Observe that
$W \subseteq (W \oplus I_{\sigma (W)})/\delta (\sigma (W))$ and by
\ref{4}.
$I_{\sigma (W)} \subseteq (W \oplus I_{\sigma (W)})/\delta (\sigma (W))
\simeq W/\sigma (W) \oplus I_{\sigma (W)}$. Since $x \in \sigma (W)$
maps to $(x,0)+\delta(\sigma(W)) = (0,-x)+\delta(\sigma(W))$, which is in
the image of $I_{\sigma (W)}$, it is possible to pass to the quotient with
respect to $W/\sigma (W)$ and $W \hookrightarrow I_{\sigma (W)}$.

$5. \Rightarrow 1.$ The last observation holds especially for all maximal
elements in $F_{\sigma (W)}$; $I_{\sigma (W)}$ is a greatest element.

Property \ref{2} is immediately verified by constructing an upper bound as a
quotient of the direct sum of all comodules of a totally ordered subset, i.e.
by forming its direct limit.
$\Box $

The proof of Lemma \ref{repint} is basically category theoretical
and relies on the facts that comodules always have a socle and that factor
comodules exist. In the language of category theory a comodule with the
property \ref{4}, i.e. a comodule that splits from every comodule that
contains it as a subcomodule, is called an injective. Clearly any direct
summand of an injective is also an injective. Using this terminology and
given a full set of simple $C$-comodules $F_0$ choose a corresponding set of
injective covers $F$, that is a set of indecomposable injective comodules
the socles of which are in $F_0$.

For every comodule $\beta :V \to V \otimes C$ there is a coefficient mapping
\begin{equation}
\phi_V : V^* \otimes V \to C
\end{equation}
by right linear extension of forms,
$\phi_V (\omega \otimes v) = \omega (\beta (v))$.
It follows that
\begin{equation}
(\phi_V \otimes id) \circ (id \otimes \beta ) = \Delta \circ \phi_V;
\end{equation}
hence $\phi_V $ is a homomorphism of (right) comodules. From Lemma
\ref{repint}, $C$ is spanned by the coefficients of the full set of
indecomposable injectives $F$:
\begin{equation}\label{coeffsum}
C = \sum_{I \in F} \phi_I (I^* \otimes I).
\end{equation}
Under favourable circumstances, e.g. if $C$ is semisimple, there is a
subfamily $F' \subseteq F$ such that
$ C = \oplus_{I \in F'} \phi_I (I^* \otimes I)$. If furthermore
$ker(\phi_I) = \lbrace 0 \rbrace$ for all $I \in F'$ this
decomposition of $C$ is the Peter Weyl lemma.

Let $\sigma(I)=I_0$. Then there is a subfamily $F' \subseteq F$ such that
\begin{equation}\label{sigmac}
\sigma(C) = \sum_{I \in F} \phi_I (I_0^* \otimes I_0)
\simeq \bigoplus_{I \in F'} m_I I_0.
\end{equation}
Due to the grading, the multiplicities $m_I$ do in general not equal
$dim(I_0)$. But, via $\phi_I$, every copy of $I_0$ in $C$ is contained in
a copy of its injective cover. From (\ref{coeffsum}) and Lemma \ref{repint}
it follows that
\begin{equation}
C \simeq \bigoplus_{I \in F'} m_I I \oplus X,
\end{equation}
where $X$ is a subcomodule of $C$. Since $\sigma(X) = \lbrace 0 \rbrace$ by
(\ref{sigmac}),
\begin{equation}\label{costruc}
C \simeq \bigoplus_{I \in F'} m_I I.
\end{equation}
This is Green's structure theorem for coalgebras. Especially $C$ itself is
injective.

With the paradigm of compact Lie groups in mind this result can be seen as
a decomposition of $C$ in a direct sum of principal vector spaces of,
loosely spoken, the center of the dual algebra $C^*$. As a corollary for
instance  there
is always a well structured theory of Fourier analysis on compact Lie
supergroups or quantisations thereof. The failure of the Peter Weyl lemma
for these may be looked at as the non diagonalisability of their algebras of
Casimir operators, the indecomposable injectives playing the role of
simultaneous principal vector spaces. These are indecomposable as comodules
but clearly decomposable as principal vector spaces of a given Casimir
operator.

Up to this point the implications are insensitive to any notion of a grading
as long as the properties of the category of comodules that were essential
in the proof of Lemma \ref{repint} are not lost. Based on Schur's lemma
and Burnside's theorem \cite{Ber:87} \cite{Sch:84} it
is possible to be slightly more explicit about this decomposition in the
present $Z_2$-graded context when working over an algebraically closed field.
In this case the commutant of a simple comodule is either one or two
dimensional. Its even part is proportional to the identity and its odd part
is spanned by a square root of minus the identity. The odd part is nonzero if
and only if the comodule is reducible in the nongraded sense. Hence the space
of coefficients of a reducible graded simple comodule faces additional
constraints.

For a comodule $W$ let $W'$ be $W$ with its grading reversed. Consider the
simple $C$-comodules modulo their grading and fix a family $\hat F$ of
indecomposa\-ble injective covers with a distinguished grading such that each
appears as a subcomodule of $C$. Let $\hat F_1$ be the subfamily of injective
covers of absolutely irreducibles in $\hat F = \hat F_1 \dot \cup \hat F_2$.
An inspection of Burnside's theorem \cite{Ber:87} yields

\begin{proposition}
A $Z_2$-graded coalgebra $C$ over an algebraically closed field decomposes as
\begin{equation}
C \simeq \bigoplus_{I \in \hat F_1} (dim( \sigma (I)_0) I \oplus
dim( \sigma (I)_1) I') \oplus
\bigoplus_{I \in \hat F_2} (\frac{1}{2} dim( \sigma (I)) I).
\end{equation}
\end{proposition}

This decomposition has an interesting consequence. Since every comodule is
contained in a direct sum of copies of $C$, indecomposable injectives come
with precisely two distinct gradings, which are reverse to each other.

A key role in the investigation of the structure of a given coalgebra is
played by induced comodules. Suppose $D$ is another coalgebra with
comultiplication $\Delta'$ and counit $\varepsilon'$ such that
$\pi : C \to D $ is a surjective mapping of coalgebras, i.e. $\pi $ is onto,
$\Delta' \circ \pi = (\pi \otimes \pi) \circ \Delta$ and $\varepsilon' \circ
\pi = \varepsilon$. For a $D$-comodule $\beta :V \to V \otimes D$ let
\begin{equation}
ind_D^C(V) = \lbrace x \in V \otimes C \vert (\beta \otimes id)(x) =
(id \otimes (\pi \otimes id)\Delta )(x) \rbrace.
\end{equation}
$ind_D^C(V)$ is a $C$-comodule with structure map
$\delta : ind_D^C(V) \to ind_D^C(V) \otimes C$,
$\delta = (id \otimes \Delta)\vert_{ind_D^C(V)}$. It is straightforward to
see that $\delta $ is well defined. In the present context it is important to
understand that $C \simeq ind_D^C(D)$ as $C$-comodules. To find this consider
the mapping $(\pi \otimes id)\Delta : C \to D \otimes C$. It is again
straightforward to show that the image of $(\pi \otimes id)\Delta $ sits
within $ind_D^C(D)$. Due to $\varepsilon' \circ \pi = \varepsilon$,
$\cdot (\varepsilon' \otimes id)$ is a left inverse to
$(\pi \otimes id)\Delta $. Hence $C \hookrightarrow ind_D^C(D)$. To show that
this mapping is onto as well let $\lbrace d_j \rbrace_{j \in J} \subset D$ be
a basis and $\Delta'(d_k) = \sum_{j \in J} d_j \otimes D^j_k$ for all
$k \in J$. Then $x = \sum_{j \in J} d_j \otimes x^j \in ind_D^C(D)$ is
equivalent to
$(\pi \otimes id)\Delta (x^k) = \sum_{j \in J} D^k_j \otimes x^j$ for all
$k \in J$. Consequently
\begin{eqnarray}
(\pi \otimes id)\Delta \cdot (\varepsilon' \otimes id)(x)
& = & (\pi \otimes id)\Delta (\sum_{j \in J} \varepsilon'(d_j)x^j) \nonumber \\
= \sum_{j \in J} \varepsilon'(d_j) (\sum_{k \in J} D^j_k \otimes x^k)
& = & \sum_{k \in J} (\sum_{j \in J} \varepsilon'(d_j) D^j_k) \otimes x^k
= x. \nonumber
\end{eqnarray}
Hence $\cdot (\varepsilon' \otimes id)$ inverts $(\pi \otimes id)\Delta $ and
$C \simeq ind_D^C(D)$. If the condition
$\varepsilon' \circ \pi = \varepsilon$ were dropped,
$(\pi \otimes id)\Delta : C \to ind_D^C(D)$ would still be onto but in
general not injective.

Now any decomposition of $D$, especially that into indecomposable injectives
$D \simeq \bigoplus_{\nu } m_{\nu } I_{\nu }$, yields a decomposition of $C$
as
\begin{equation}
C \simeq \bigoplus_{\nu } m_{\nu } \ ind_D^C(I_{\nu }).
\end{equation}
Since direct summands of injectives are obviously injectives,
this implies in particular that induction carries injectives to injectives.
The study of induced modules forms the basic task in the investigation of the
structure of the representation theory of coalgebras. This task may be a very
complex one in general. Even in the case of Lie superalgebras and related
coalgebras it is therefore advisable to consider examples that come with
additional structures within their induced modules. This happens for type I
Lie superalgebras where the additional structures are filtrations by means of
Kac modules.

\section{The general linear supergroup}
The general linear supergroup offers an interesting example for an
illustration of the general theory. Induction from the base Lie group comes
with special filtrations that allow for some insight into the substructure
of the corresponding coalgebra. From now on all constructions are meant to
live over the base field of complex numbers unless otherwise mentioned.

It is sensible to begin with polynomials. Let $C(p,q)$ be the bialgebra
that is generated by a supermatrix $((c^i_j))$ of dimension
$(p+q) \times (p+q)$. The multiplication is meant to be graded commutative
and the comultiplication comes from ordinary matrix multiplication as
$\Delta (c^i_j) = \sum_k c^i_k \otimes c^k_j$. The counit is defined from
$\varepsilon (c^i_j) = \delta ^i_j$.

At the same time $((c^i_j))$ defines a comodule structure
$\beta :V \to V \otimes C(p,q)$ on a $p+q$ dimensional graded vector space
with respect to a basis
$\lbrace e_i \rbrace$ by $\beta (e_i) = \sum_j e_j \otimes c^j_i$,
the defining representation. Consequently the $k$-homogeneous part
$C(p,q)_k$ of $C(p,q)$ is spanned by the coefficients of the comodule
$V^{\otimes k}$. Especially $C(p,q) \simeq \bigoplus_{k \ge 0} C(p,q)_k$
is a decomposition into finite dimensional coalgebras. The transposition
$x \otimes y \mapsto (-1)^{|x||y|}y \otimes x$, for homogeneous elements
$x \in V_{|x|}$, $y \in V_{|y|}$, induces a representation of the
permutation group $\Sigma_k$ of $k$ objects on $V^{\otimes k}$. Its
enveloping algebra is the commutant of $C(p,q)_k^*$. Since the endomorphism
ring of a semisimple module is semisimple this implies that $C(p,q)_k$
is semisimple if the characteristic of the base field is either zero or
exceeds $\#(\Sigma_k)=k!$. This is in complete analogy to the case of
polynomial representations of $GL_n$ \cite{Gre:81}. Furthermore, in
characteristic zero, $C(p,q)$ is semisimple. This theory of polynomial
representations of the general linear superalgebra is worked out
in detail in \cite{Mui:91}. It is worth to remark that the structure of this
argument still holds for one parameter quantisations of $C(p,q)$. The major
change is that the commutant of say the $k$-homogeneous part depends on
the deformation parameter. If this dependence is reflected in defining
relations that are polynomial in the parameter, inspection of its Killing
form shows that the breakdown of semisimplicity is governed by zeros of
polynomial equations in the deformation parameter. Hence one finds
semisimplicity for polynomial representations of degree $k$ in
characteristic zero or greater than $k!$ for all but finitely many values
of the deformation parameter. Finally in characteristic zero a one parameter
quantisation of $C(p,q)$ is generically semisimple apart from countably many
values of the deformation parameter.

To pass to a supergroup, $C(p,q)$ has to be enlarged to a Hopf algebra. To
see how this can be done the inversion of supermatrices has to be considered.
With respect to the natural block decomposition induced by the grading of
$V \simeq V_0 \oplus V_1$ the formal inverse of
$((c^i_j)) =
{\footnotesize \left (\begin{array}{cc} A & B \\ C & D \end{array} \right)}$
reads
\begin{equation}
\left (\begin{array}{cc} A & B \\ C & D \end{array} \right)^{-1} =
\left (\begin{array}{cc} (A - BD^{-1}C)^{-1} & -A^{-1}B (D - CA^{-1}B)^{-1}
\\ -D^{-1}C (A - BD^{-1}C)^{-1} & (D - CA^{-1}B)^{-1} \end{array} \right).
\end{equation}
It follows that the matrix of generators $((c^i_j))$
is invertible as soon as $det(A)$ and $det(D)$ are. To get a Hopf algebra
$A(p,q)$ with involutive antipode, $C(p,q)$ has to be localized at the monoid
which is generated by $det(A)$ and $det(D)$.
For notational convenience let
$((c^i_j))^{-1} =
{\footnotesize \left (\begin{array}{cc} \tilde A & \tilde B \\
\tilde C & \tilde D \end{array} \right)}$.

The structure of $A(p,q)$ can be analysed with the help of induced modules
with respect to the coalgebra $A_0(p,q) \simeq A(p,0) \otimes A(0,q)$. The
matrix of generators of $A(p,0)$ will be denoted by $A_0$, those
of $A(0,q)$ by $D_0$. The projection $\pi : A(p,q) \to A_0(p,q)$ is defined
componentwise by $\pi (A) = A_0$,
$\pi (D) = D_0$, $\pi (B) = 0$, $\pi (C) = 0$. Let
$\beta :V \to V \otimes A_0(p,q)$ be a $A_0(p,q)$-comodule
and $\lbrace e_i \rbrace \subset V$ a basis such that
$\beta (e_i) = \sum_j e_j \otimes {\cal D}(A_0,D_0)^j_i$.
Now $x = \sum_i e_i \otimes x^i \in ind_{A_0(p,q)}^{A(p,q)}(V)$ means
$(\pi \otimes id)\Delta (x^i) = \sum_j {\cal D}(A_0,D_0)^i_j \otimes x^j$.
This equation is straightforward to solve with the result
$x^i(A,B,C,D) = \sum_j {\cal D}(A,D)^i_j F^j(A^{-1}B,D^{-1}C)$. Hence the
elements of $ind_{A_0(p,q)}^{A(p,q)}(V)$ are parameterised as
\begin{equation}
x(A,B,C,D) = \sum_{i,j} e_i \otimes {\cal D}(A,D)^i_j F^j(A^{-1}B,D^{-1}C)
\end{equation}
and $dim(ind_{A_0(p,q)}^{A(p,q)}(V))=2^{2pq}dim(V)$.
The explicit realisation of the structure map
\begin{equation}
\delta_V : ind_{A_0(p,q)}^{A(p,q)}(V) \to
ind_{A_0(p,q)}^{A(p,q)}(V) \otimes A(p,q)
\end{equation}
reads
\begin{equation}
\delta_V (x) = \sum_{i,j,k} e_i \otimes {\cal D}(A_1,D_1)^i_j
{\cal D}( A_2 + \xi_1 C_2 , D_2 + \eta_1 B_2 )^j_k \cdot
\Delta (F^k(\xi ,\eta ))
\end{equation}
with $\xi = A^{-1}B$, $\eta = D^{-1}C$ and
$\Delta (\xi ) = (A_2 + \xi_1 C_2)^{-1}( B_2 + \xi_1 D_2 )$,
$\Delta (\eta ) = (D_2 + \eta_1 B_2)^{-1}(C_2 + \eta_1 A_2)$. For the
convenience of notation the first and second factor in the tensor product
after application of the comultiplication are distinguished by subscripts
and identity matrices are not written explicitly.

There are two possibilities for similarity transformations, motivated by
either distinguishing the variables $\xi$ or $\eta$ respectively. These arise
from the identities
\begin{equation}
\Delta (A-BD^{-1}C) = \Delta (\tilde A^{-1}) =
\tilde A_1^{-1}(\tilde A_2 - \tilde B_2 \eta_1)^{-1}
\end{equation}
or
\begin{equation}
\Delta (D-CA^{-1}B) = \Delta (\tilde D^{-1}) =
\tilde D_1^{-1}(\tilde D_2 - \tilde C_2 \xi_1)^{-1}
\end{equation}
Parameterising
\begin{equation}
x(A,B,C,D) = \sum_{i,j} e_i \otimes
{\cal D}(A,\tilde D^{-1})^i_j \ x_1^j(\xi,\eta),
\end{equation}
the comodule structure is represented by
\begin{equation}
\delta_V(x(A,B,C,D)) = \sum_{i,j} e_i \otimes
{\cal D}(A_1,\tilde D_1^{-1})^i_j \ \delta^1_V(x_1^j(\xi,\eta))
\end{equation}
with
\begin{equation}
\delta^1_V(x_1^j(\xi,\eta)) = \sum_k
{\cal D}(A_2+\xi_1C_2,(\tilde D_2 - \tilde C_2 \xi_1)^{-1})^i_k
\Delta (x_1^k(\xi,\eta)).
\end{equation}
Correspondingly for
\begin{equation}
x(A,B,C,D) = \sum_{i,j} e_i \otimes
{\cal D}(\tilde A^{-1},D)^i_j \ x_2^j(\xi,\eta)
\end{equation}
it follows that
\begin{equation}
\delta_V(x(A,B,C,D)) = \sum_{i,j} e_i \otimes
{\cal D}(\tilde A_1^{-1},D_1)^i_j \ \delta^2_V(x_2^j(\xi,\eta))
\end{equation}
with
\begin{equation}
\delta^2_V(x_2^j(\xi,\eta)) = \sum_k
{\cal D}((\tilde A_2 - \tilde B_2 \eta_1)^{-1},D_2+\eta_1B_2)^j_k
\Delta (x_2^k(\xi,\eta)).
\end{equation}

The substructure of these comodules is partly accessible by looking at
suitable analogues of parabolic subgroups. The grading itself suggests a
natural choice that was used by Kac in his original work on the finite
representations of Lie superalgebras. To see how this looks in the present
context one has to consider coalgebras generated by either the upper or
lower block triangular entries of the matrix of generators. Hence
there are coalgebras $A_{+/-}(p,q)$ with projections
$\pi_{+/-}:A(p,q) \to A_{+/-}(p,q)$ given by
$\pi_{+/-} (A) = A_{+/-}$, $\pi_{+/-} (D) = D_{+/-}$, $\pi_+ (B) = B_+$,
$\pi_- (B) = 0$, $\pi_+ (C) = 0$, $\pi_- (C) = C_-$. These allow for the
notion of primitive elements. Let $\gamma : W \to W \otimes A(p,q)$ be
a comodule. $x \in W$ is called primitive if
$(id \otimes \pi_+)\gamma (x) \in W \otimes A_+(p,q)$
is independent of $B_+$, antiprimitive if
$(id \otimes \pi_-)\gamma (x) \in W \otimes A_-(p,q)$
is independent of $C_-$.

In order to find the (anti)primitive elements in the induced modules consider
\begin{equation}
(id \otimes \pi_+)\delta^1_V(x_1^j(\xi,\eta)) =
\sum_k {\cal D}(A_2,D_2)^i_k \ (id \otimes \pi_+)\Delta (x_1^k(\xi,\eta))
\end{equation}
and
\begin{equation}
(id \otimes \pi_-)\delta^2_V(x_2^j(\xi,\eta)) =
\sum_k {\cal D}(A_2,D_2)^j_k \ (id \otimes \pi_-)\Delta (x_2^k(\xi,\eta)).
\end{equation}
It remains to solve the equations
\begin{eqnarray}
(id \otimes \pi_+)\Delta (f_+(\xi ,\eta )) & = &
f_+(A_{+2}^{-1}( B_{+2} + \xi_1 D_{+2}),
(D_{+2} + \eta_1 B_{+2})^{-1}\eta_1 A_{+2}) \nonumber \\
& = &
f_+(A_{+2}^{-1}\xi_1 D_{+2},D_{+2}^{-1}\eta_1 A_{+2})
\end{eqnarray}
and
\begin{eqnarray}
(id \otimes \pi_-)\Delta (f_-(\xi ,\eta )) & = &
f_-((A_{-2} + \xi_1 C_{-2})^{-1}\xi_1 D_{-2},
D_{-2}^{-1}(C_{-2} + \eta_1 A_{-2})) \nonumber \\
& = &
f_-(A_{-2}^{-1}\xi_1 D_{-2},D_{-2}^{-1}\eta_1 A_{-2}).
\end{eqnarray}
The result is immediate from the formal substitution of
$B_{+2} = - \xi_1 D_{+2}$ or $C_{-2} = - \eta_1 A_{-2}$,
since the corresponding equations do not depend on these variables:
\begin{equation}
f_+(\xi ,\eta ) = f_+(0,(1-\eta \xi )^{-1}\eta ) \quad \mbox{and} \quad
f_-(\xi ,\eta ) = f_-((1-\xi \eta )^{-1}\xi ,0).
\end{equation}
Finally a general primitive element of $ind_{A_0(p,q)}^{A(p,q)}(V)$ has the
form
\begin{equation}
x_+(A,B,C,D) = \sum_{i,j} e_i \otimes {\cal D}(A,\tilde D^{-1})^i_j \
F_+^j((1-\eta \xi )^{-1}\eta )
\end{equation}
and a general antiprimitive element correspondingly
\begin{equation}
x_-(A,B,C,D) = \sum_{i,j} e_i \otimes {\cal D}(\tilde A^{-1},D)^i_j \
F_-^j((1-\xi \eta )^{-1}\xi ).
\end{equation}

A short calculation establishes the relations
\begin{eqnarray} \lefteqn{
\Delta ((1-\xi \eta )^{-1}\xi ) =} \\ & &
\tilde A_2 (1+B_2D_2^{-1}\eta_1)
\lbrace B_2D_2^{-1}\ +\ (1-\xi_1\eta_1)^{-1}\xi_1 \
(1+\eta_1B_2D_2^{-1})\rbrace D_2, \nonumber
\end{eqnarray}
\begin{eqnarray} \label{filt} \lefteqn{
\Delta ((1-\eta \xi )^{-1}\eta ) =} \\ & &
\tilde D_2 (1+C_2A_2^{-1}\xi_1)
\lbrace C_2A_2^{-1}\ +\ (1-\eta_1\xi_1)^{-1}\eta_1 \
(1+\xi_1C_2A_2^{-1})\rbrace A_2. \nonumber
\end{eqnarray}
Hence, using the parameterisation
\begin{equation}
x(A,B,C,D) = \sum_{i,j} e_i \otimes {\cal D}(A,\tilde D^{-1})^i_j \
x_1^j(\xi ,(1-\eta \xi )^{-1}\eta ),
\end{equation}
it follows from equation (\ref{filt}) that
a filtration of $ind_{A_0(p,q)}^{A(p,q)}(V)$ is inherited from the expansion
with respect to the composite variables $(1-\eta \xi )^{-1}\eta $. That means
for every $j \le pq$ let $W_j(V)$ be the subcomodule of
$ind_{A_0(p,q)}^{A(p,q)}(V)$ spanned by all elements with $x_1^k$ of degree
less or equal than $j$ with respect to the entries of
$(1-\eta \xi )^{-1}\eta $. Consequently
\begin{equation}
\lbrace 0 \rbrace \subset W_0(V) \subset W_1(V) \subset \dots \subset
W_{pq}(V) = ind_{A_0(p,q)}^{A(p,q)}(V)
\end{equation}
and from the
above relations $W_j(V)/W_{j-1}(V) \simeq W_0(V \otimes V_j)$ with
$V_j$ the $A_0(p,q)$-comodule
spanned by the j-th order monomials of $(1-\eta \xi )^{-1}\eta $. The
invariant subspace $W_0(V)$, corresponding to degree zero in
$(1-\eta \xi )^{-1}\eta $, is spanned by elements of the form
\begin{equation}
x(A,B,C,D) = \sum_{i,j} e_i \otimes {\cal D}(A,\tilde D^{-1})^i_j \
x_1^j(\xi ).
\end{equation}
The only primitive vectors contained in $W_0(V)$ are those with
$x_1^k(\xi)$ independent of $\xi$. Obvious antiprimitive vectors in these
representations are those with all $x_1^k$ proportional to the highest power
in the entries of $\xi $; however it may happen that there are more
of them, implying that $W_0(V)$ may be reducible but indecomposable. These
comodules correspond to one type of Berezin's elementary representations
\cite{Ber:87} which are in turn equivalent to lowest weight Kac modules
\cite{Kac:77b}, \cite{Kac:78}. The irreducible ones correspond to typical
irreducible modules in Kac's terminology.

Analogously
\begin{equation}
x(A,B,C,D) = \sum_{i,j} e_i \otimes {\cal D}(\tilde A^{-1},D)^i_j \
x_2^j((1-\xi \eta )^{-1}\xi ,\eta )
\end{equation}
leads to a filtration of $ind_{A_0(p,q)}^{A(p,q)}(V)$ coming from the
expansion with respect to the variables $(1-\xi \eta )^{-1}\xi $. The
invariant subspace corresponding to degree zero is in this case spanned by
elements of the form
\begin{equation}
x(A,B,C,D) = \sum_{i,j} e_i \otimes {\cal D}(\tilde A^{-1},D)^i_j \
x_2^k(\eta ).
\end{equation}
These comodules correspond to the second type of Berezins elementary
representations which are basically highest weight Kac modules. The only
antiprimitive elements contained are those with $x_2^k(\eta)$ independent of
$\eta$. There are primitive elements with all $x_2^k$ proportional to the
highest power in the entries of $\eta $ and the comodules correspond to
highest weight Kac modules of the general linear Lie superalgebra. Again
these representations are typical in the sense of Kac if irreducible.

To conclude this section, let $V \subset A_0(p,q)$ be a subcomodule with its
basis elements $e_i = E_i(A_0,D_0)$. Then
$$
\sum_i \varepsilon'(e_i) {\cal D}(X,Y)^i_j = \sum_i E_i(1,1)
{\cal D}(X,Y)^i_j = E_j(X,Y)
$$
and as elements of $A(p,q)$ the above choices of coordinates read either
\begin{equation}
x(A,B,C,D) = \sum_{i} E_i(A,D(1-\eta \xi)) \
x_1^i(\xi ,(1-\eta \xi )^{-1}\eta )
\end{equation}
or
\begin{equation}\label{para}
x(A,B,C,D) = \sum_{i} E_i(A(1-\xi \eta),D) \
x_2^i((1-\xi \eta )^{-1}\xi ,\eta )
\end{equation}
since $\tilde A^{-1} = A(1-\xi \eta)$ and $\tilde D^{-1} = D(1-\eta \xi)$.
Here $x$ is abused for $\cdot (\varepsilon' \otimes id) x$. The explicit
splitting into indecomposable injectives has to be derived from the
information about the (anti)primitive elements and the associated
filtrations, which is still a formidable task when considered in full
generality. Introducing weight spaces and using Kac's results about Casimir
elements it is easy to verify, that the elementary invariant subspaces
$W_0(V)$ split from $ind_{A_0(p,q)}^{A(p,q)}(V)$ iff they are typical
irreducible. However, in the atypical cases the corresponding indecomposable
injectives are strictly larger than these submodules as is to be illustrated
below.

\section{Examples}
Finally the general theory is to be illustrated by simple examples.
Fortunately the easiest case, i.e $A(1,1)$, which describes a part of the
locally finite modules of the Lie superalgebra $gl(1,1;{\bf C})$, shows
already the general features without demanding involved computations.
Therefore it is well suited for illustrative purposes. This should at the
same time clarify confusions that arose in the literature concerning Fourier
analysis on the unitary supergroup $U(1,1)$ \cite{Guh:93}.

To begin with, the underlying bosonic coalgebra $A_0(1,1)$ is generated by
two independent bosonic variables $A_0$ and $D_0$ and their inverses. The
decomposition of $A_0(1,1)$ corresponds to its ${\bf Z} \times {\bf Z}$
grading. The simple comodules $V_{n_1,n_2}$ are one dimensional and labelled
by two integers $n_1,n_2$. As a basis of $V_{n_1,n_2}$ within $A_0(1,1)$
choose $A_0^{n_1}D_0^{n_2} \in A_0(1,1)$. With respect to the parameterization
(\ref{para}), elements of
$ind_{A_0(1,1)}^{A(1,1)}(V_{n_1,n_2}) \subset A(p,q)$ are written as
\begin{equation}
x(A,B,C,D) = \tilde A^{-n_1} D^{n_2} \ x_2((1-\xi \eta)^{-1}\xi,\eta).
\end{equation}
The primitive vectors in this representation are given by
\begin{eqnarray}
x_+(A,B,C,D) & = & A^{n_1}\tilde D^{-n_2} \ F_+((1-\eta\xi)^{-1}\eta) \\
& = & \tilde A^{-n_1} D^{n_2} \ (1-\eta\xi)^{n_1+n_2} F_+(\eta).
\end{eqnarray}
Obviously the cases $n_1+n_2 = 0$ bear a significant difference compared to
$n_1+n_2 \not= 0$; all primitive elements live inside the elementary
subcomodule which as a consequence cannot split from the induced
representation. Hence $n_1+n_2 = 0$ marks four dimensional injective modules
that extend one dimensional irreducible representations, which are atypical in
Kac's terminology. Their Jordan H\"older sequences have length four. In the
case of $n_1+n_2 \not= 0$ the comodule $ind_{A_0(1,1)}^{A(1,1)}(V_{n_1,n_2})$
splits into the direct sum of two typical irreducible representations spanned
by either the elements
\begin{eqnarray}
y_1^{n_1,n_2}(A,B,C,D) & = & \tilde A^{-n_1} D^{n_2} \cdot \eta, \\
y_2^{n_1,n_2}(A,B,C,D) & = & \tilde A^{-n_1} D^{n_2} \cdot 1
\end{eqnarray}
or
\begin{eqnarray}
y_3^{n_1,n_2}(A,B,C,D) & = & \tilde A^{-n_1} D^{n_2} \cdot
(1-\eta\xi)^{n_1+n_2} \\
& = & \tilde A^{-n_1} D^{n_2} \cdot (1-(n_1+n_2)\eta\xi), \nonumber \\
y_4^{n_1,n_2}(A,B,C,D) & = & \tilde A^{-n_1} D^{n_2} \cdot \xi
\end{eqnarray}
respectively. In the case of $n_1+n_2 = 0$ the basis vector $y_3$ may be
replaced by
\begin{equation}
{y_3^{n_1,n_2}}'(A,B,C,D) = \tilde A^{-n_1} D^{n_2} \cdot \eta\xi.
\end{equation}

This completes the explicit construction of the Fourier analysis of $A(1,1)$
and, by specialisation, that of the compact unitary supergroup $U(1,1)$. It is
easy to verify explicitly, that the standard quadratic Casimir element of
$U(gl(1,1;{\bf C}))$ is proportional to the identity on the two dimensional
injectives corresponding to $n_1+n_2 \not= 0$ and falls into two
one dimensional and one two dimensional Jordan blocks when acting on one of the
four dimensional injectives with $n_1+n_2 = 0$. The latter are all degenerate
and form principal vector spaces that extend the kernel of the Casimir
operator.

The next step should be an investigation of $A(2,1)$. Since a lot has been
published about the Lie superalgebra $gl(2,1;{\bf C})$ or $sl(2,1;{\bf C})$
respectively this case will be treated by an informal discussion of its
features, relating it to the literature. The structure of the induced modules
is slightly more involved than in the $A(1,1)$ example. In terms of
elementary representations the Kac filtrations of the induced comodules have
four composition factors. The introduction of weights and inspection of those
of the primitive elements shows that reducible elementary representations
always occur in pairs as composition factors of induced representations. They
themselves have two simple composition factors. This leads to a picture that
is very similar to the results concerning $A(1,1)$. Indecomposable
injectives are either typical irreducible representations or they extend
atypical irreducible representations and then have Kac filtrations of length
two and Jordan Hoelder sequences of length four. These results are contained
in reference \cite{Su:92}, where they were derived in a Lie superalgebra
context, without realising that the matter of investigation was the coalgebra
encoding the locally finite representations.

\section{Conclusion}

A discussion of the representation theory of graded coalgebras was given
which especially sheds some light on the role of atypical irreducible
representations of type I Lie superalgebras. The basic problem for further
advances is the classification of their injectives. It would be interesting
if that could be carried out at least for all $A(p,q)$. Partial results
come from the construction of the (anti)primitive elements in induced
representations. Others can be deduced from Kac's original work on the
characters of simple Lie superalgebras, especially to obtain information
about the blocks \cite{Gre:76} of e.g. $A(p,q)$. One might speculate if the
injective comodules of $A(p,q)$ could be be characterised by the property to
have Kac filtrations by means of both types of elementary representations
as it is in the case of $A(1,1)$  and $A(2,1)$. If this is not the case, a
counterexample would be of interest. I hope to come back to these questions
once the details piece together to a satisfying additional insight.

{\bf Acknowledgements}. I would like to thank Professor Zirnbauer for
discussions as well as Dr. N. Andruskiewitsch, who furthermore brought
reference \cite{Su:92} to my attention. Professor Donkin pointed references
\cite{Gre:76},  \cite{Gre:81} and \cite{Mui:91} out to me. This work had its
origin within the SFB 341 K\"oln Aachen J\"ulich and was completed under
support by the EC.

\end{document}